\documentclass[runningheads]{llncs}
\usepackage[T1]{fontenc}
\usepackage{array}
\usepackage{graphicx}
\usepackage{amsmath}
\usepackage{amssymb}

\usepackage{listings}
\usepackage{xcolor}

\definecolor{mycolor}{rgb}{0.0, 0.5, 0.0}
\begin{document}
\title{QIris: Quantum Implementation of Rainbow Table Attacks}
%
%

\author{Lee Jun Quan\inst{1} \and
Tan Jia Ye\inst{1} \and
Goh Geok Ling\inst{1} \and
Vivek Balachandran\inst{2}}

\institute{
\email{\{2201509, 2201862, 2202614\}@sit.singaporetech.edu.sg}\\
\and
\email{vivek.b@singaporetech.edu.sg}\\
Singapore Institute of Technology}

%
%
\maketitle           
\begin{abstract}
This paper explores the use of Grover's Algorithm in the classical rainbow table, uncovering the potential of integrating quantum computing techniques with conventional cryptographic methods to develop a Quantum Rainbow Table. It leverages on quantum concepts and algorithms, including the principles of qubit superposition, entanglement, and teleportation, coupled with Grover's Algorithm, to enable a more efficient search through the rainbow table. The paper also details the current hardware constraints and a workaround to produce better results in the implementation stages. Through this work, we develop a working prototype of the quantum rainbow table and demonstrate how quantum computing could significantly improve the speed of cyber tools such as password crackers and thus impacting the cybersecurity landscape. 

\end{abstract}

\section{INTRODUCTION}
In this paper, we explore the potential of using quantum computing to improve the speed of cracking hashes in legacy systems, which opens to new possibilities, highlighting the need for quantum-resilient cryptographic techniques. This research aims to create a Quantum Rainbow Table, merging and applying principles of quantum mechanics together with Grover’s Algorithm \cite{younes2008strength}. This exhibits the abilities of quantum-assisted search techniques, placing emphasis on the impact quantum computing has on cryptographic analysis. The goal is to explore potential implications of quantum computing on cryptography, with the aim of contributing to advancements in cybersecurity practices. 

\subsection{Background}
Safeguarding of password data is critical in the cybersecurity landscape. In most systems, passwords are stored as cryptographic hash values and not in plaintext, in order to prevent attackers from accessing such confidential information. However, if the attackers manage to obtain this information, they would then be able to devise methods to compare the hashes using a precomputed table, known as a Rainbow Table. These tables store the output of values put through a hash function, which then enables attackers to conduct rapid searching of said hashes in order to decipher encrypted passwords \cite{oechslin2003making} in an efficient and effective manner \cite{kumar2013rainbow}. Our work explores the potential of using the prowess of quantum computing in order to expand on this efficiency in a manner greater than that of classical search algorithms.

\subsection{Motivations}
Classical rainbow tables typically involve a linear search through the table,  resulting in a worst-case time complexity of $O(N)$, where N is defined as the number of entries in the table. On the other hand, by leveraging Grover's Algorithm, a quantum algorithm designed for searching unsorted tables, the time complexity for the searches can be significantly reduced. In comparison to classical methods \cite{khurana2023implementation}, Grover’s Algorithm runs with a time complexity of  $O(\sqrt{N})$ iterations, which is a quadratic improvement in the time taken for the search process. However, searching using Grover’s Algorithm will indicate if the target entry exists within the database but it does not return the target's position or identity. Hence, in order to determine the exact position and identity of the target entry, additional steps are needed \cite{younes2008strength}.  

To the best of our knowledge, no practical quantum solutions have been explored for rainbow table searches with MD5 hashes. However, a study identifying the quantum resources required for applying Grover’s Algorithm to extract keys from a small number of AES plaintext-ciphertext pairs is discussed in \cite{grassl2016applying}. 

In recent years, quantum properties have been increasingly leveraged to enhance classical cybersecurity implementations, including works in opaque quantum predicate used in hybrid quantum-classical systems with a high level of accuracy \cite{balachandran2021quantum}, development of a quantum key distribution protocol \cite{bennett1992quantum}, and a quantum implementation of AES \cite{app11199085}. However, in the domain of rainbow tables, the closest work that we could find was a theoretical work on Hellman's time-memory trade-off on the Rainbow Table variant \cite{dunkelman2024quantum}. These are the motivations for this work to develop a working prototype of the quantum rainbow table, providing a practical implementation to validate theoretical works in this space. 

\subsection{Proposed Approach}
In QIris, MD5 hashing was selected to showcase the prototype's feasibility. The implementation is flexible and can be adapted to work with other hashing algorithms with minor code modification, which will be explained and indicated in the subsection on Rainbow Table Generation. This hybrid quantum-classical approach combines Grover’s Algorithm with classical pre-processing checks on buckets for matches. Grover’s Algorithm helps narrow down the search space by identifying the bucket, followed by performing a classical check on the bucket to determine the specific match. This hybrid approach exhibits the quantum speedup for search as compared to linear search with a worst-case time complexity of $O(N)$, ensuring that QIris can find and identify the correct plaintext in a quantum-efficient manner. QIris was developed using Python, utilizing various packages and libraries. These include Qiskit (version 1.1.0) with modules such as quantum\_info, transpiler, and circuit, Qiskit Aer (version 0.14.2) for simulation with the Aer library, along with numpy (version 1.26.4) and tweedledum (version 1.1.1) to support the necessary computations and quantum operations.

\subsubsection{Limitations and Assumptions} 
Throughout the development of QIris, we faced substantial hardware and processing constraints, resulting in a 4-qubit implementation of the Quantum Rainbow Table. To adhere to the constraints of quantum circuits, it is necessary to employ a simplified alternative hashing method \cite{pearson1990fast}, which leads to a higher hash collision. In an ideal situation without hardware constraints, the number of qubits would have no limitations, there would not be a need to implement any further hashing, and the plaintext can be represented in their ASCII binary representation. However, this would require a large number of qubits; for instance, a single character would need 8 qubits to be represented in a quantum circuit. To circumvent this limitation, keywords were mapped into 4 bit qubits representation using a simplified hash.   

In addition to hardware limitations, we worked on quantum simulators for this paper instead of real quantum computers. The quantum simulator provided an idealized version of a quantum computer on classical machines to test our work and bypasses the limitations of not having access to quantum hardware. However, this approach also means that potential quantum errors and the nuanced realities of physical quantum operations were not accounted for, as the simulators operated under the assumption of ideal quantum behavior. This assumption provided a controlled environment for algorithm development and testing, but may not fully represent the challenges encountered on actual quantum hardware.

Our implementation of the Grover’s Algorithm outputs a boolean value indicating the presence of a target in a search space (bucket). In order to obtain the position of the value in the bucket, the use of classical linear search is needed following the application of Grover’s Algorithm.

\section{METHODOLOGY}
\begin{figure}[hbt!]
    \centering
    \includegraphics[width=1\textwidth,keepaspectratio]{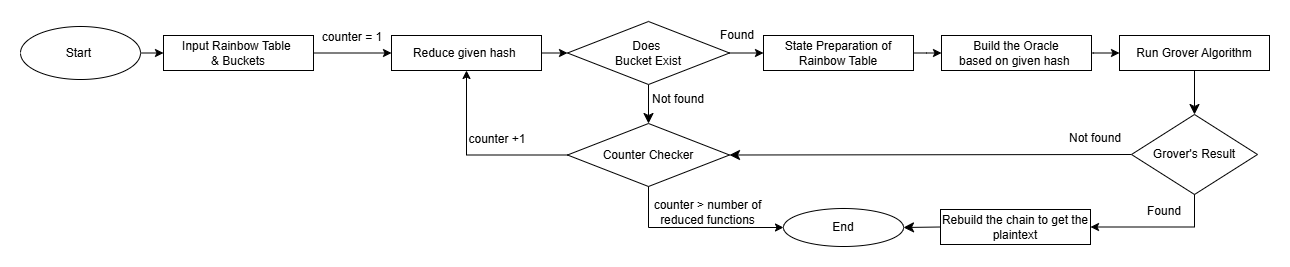}
    \caption{Methodology of QIris.}
    \label{methodology}
\end{figure}

QIris aims to leverage on the speed advantages of quantum computing to enhance the classical rainbow table approach. As described in Section 1, Grover’s Algorithm, with a time complexity of $O(\sqrt{N})$, can significantly improve search efficiency. By using Grover’s Algorithm to replace parts of the linear search process, QIris achieves a faster overall time complexity. The creation of the rainbow tables and the buckets used in QIris is a one-time process as they remain static, hence, this will not be factored in the time complexity. Meanwhile, the search for a specific hash within these tables is a dynamic process that occurs at runtime, hence, this will be factored into the time complexity, where Grover’s Algorithm is employed to accelerate the search. Figure~\ref{methodology} shows the flow of QIris during runtime. 

\subsection{Rainbow Table Generation And Buckets Creation}
\subsubsection{Rainbow Table Generation}
The rainbow table utilized in QIris is generated by pre-computing and storing both the initial chain plaintext and the final reduction plaintext into a text file format (.txt) file. To increase the table’s efficacy and diversity, four distinct reduction functions (R1, R2, R3, and R4) were used in its construction. Each reduction function converts the hash value into a plaintext of different lengths, which increases the plaintext’s diversity and raises the likelihood of a successful match. The input to the reduction function is the MD5-hashed value of the plaintext starting from the initial list in the rainbow table chain.

In our implementation of the reduction function, the first 8 characters of the hash are mapped to an integer. Each reduction function adds a unique nonce, N, to this integer (N = 2 for R1, N = 3 for R2, N = 4 for R3 and N = 1 for R4).  This integer is then converted to a plaintext by extracting numbers through modulus operation and mapping them to the base62 encoded character set ([ 0-9, a-z, A-Z ]). This process is repeated X number of times (X = 6 for R1,  X = 4 for R2, X = 5 for R3 and X = 3 for R4). The plaintext generated from R1 is hashed and acts as the input for R2. Similarly, the hashed plaintext from R2 is given as input to R3 and so on. In our implementation, the chain length is four. However, longer chain lengths can improve the performance of the rainbow table. The python code for the reduction function is shown in Listing~\ref{reductionfunction}.

\begin{lstlisting}[caption={Reduction Function},captionpos=b,label={reductionfunction}]
def reduction_function(md5_hash):
    hash_int = int(md5_hash[:8], 16) + N
    plaintext = ""
    for _ in range(X): 
        plaintext += base62_characters[hash_int % 62]
        hash_int //= 62
    return plaintext
\end{lstlisting}

\subsubsection{Bucket Creation}
QIris employs a bucket strategy to streamline the search process by organizing hashes into predefined groups, termed  “buckets”. This strategy reduces the search space that Grover’s Algorithm needs to process, thereby improving the efficiency of search through the rainbow table. This method optimizes the overall performance by ensuring that Grover’s Algorithm operates on a smaller search space, thereby accelerating the post-search process.

The final reduced plaintext is hashed using Pearson's simplified alternative hashing method \cite{pearson1990fast}, transforming it into a 16-bit integer. These integers are then stored in buckets of size 16. Specifically, integers in the range 0–15 are placed in bucket 0, those from 16–31 in bucket 1, and so forth. Once the bucket key is derived from the plaintext, the integer is inserted into the corresponding bucket using a modulo-16 operation. If a bucket corresponding to the computed bucket key does not already exist, a new bucket is created to accommodate the value. The code for the simplified alternative hashing method is shown in Listing \ref{lst:simphash} and the creation of buckets in Listing \ref{lst:buckstrat}.. 


\begin{lstlisting}[caption={Simplified Hashing},captionpos=b,label={lst:simphash}]
random.seed(44)
permutation_16bit = list(range(65535))
random.shuffle(permutation_16bit)
def hash16bit(txt):
    h = len(txt) % 65535
    for i in txt:
        h = permuation_16bit[(h + ord(i)) % 65535]
    return h
\end{lstlisting}


\begin{lstlisting}[caption={Bucket Creation},captionpos=b,label={lst:buckstrat}]
end_hashed = hash16bit(end)
bucket_key = end_hashed // 16
if bucket_key not in buckets:
     buckets[bucket_key] = []
buckets[bucket_key].append(end_hashed % 16)
\end{lstlisting}

\subsection{Iterative Implementation}
In a classical rainbow table the MD5 hash is reduced to plaintext and a linear search is conducted on a list of plaintexts at the end of rainbow table chains. In QIris, we reduce the hash to a plaintext target, $T_1$, using our reduction functions and then proceed to search it in the buckets. A classical linear search is performed to verify that the bucket key of $T_1$ exists. Grover's algorithm is then used to search for the specific target $T_1$ in the bucket. If the bucket key does not exist in the bucket list, QIris skips the Grover’s search and subsequent linear search, moving directly to the previous chain, thus saving time. After Grover’s search, if the result is found, a linear search is performed to locate for the given hash and rebuild the chain to retrieve the plaintext. If the result is not found, QIris will move on to the previous chain. This process continues until QIris has examined all four chains.  If the hash is not found in any of the chains, QIris concludes with the hash being deemed not present in the rainbow table. The code for this implementation is presented in Listing \ref{lst:intimp}.

\begin{lstlisting}[caption={Iterative  Implementation},captionpos=b,label={lst:intimp}]
funcs = get_reduction_function()
    for i in range(len(funcs)):
        i += 1
        funcs_to_use = funcs[-i:]
        remaining_funcs = funcs[:-i]
        current_hash = search_hash
        for j in range(len(funcs_to_use)):
            current_text = funcs_to_use[j](current_hash)
            if j != len(funcs_to_use) - 1:
                current_hash = md5_hash(current_text)
        current_map = hash16bit(current_text)
        bucket_key = current_map // 16
        if bucket_key not in buckets:
            continue
        good_states = buckets[bucket_key]
        lookup = current_map % 16
        if len(good_states) <= 2:
            if lookup in good_states:
                grover_result = True
            else:
                grover_result = False
        Else:
  		Grover Search Codes 
            if Grover Search Counts > 512:
                grover_result = True
            else:
                grover_result = False
        if grover_result is True:
            index = rainbow_table_end_hashed.index(current_map)
            if current_text != rainbow_table_end[index]:
                continue
            current_text = rainbow_table_start[index]
            current_hash = md5_hash(current_text)
            for k in range(len(remaining_funcs)):
                current_text = remaining_funcs[k](current_hash)
                current_hash = md5_hash(current_text)
            if current_hash == search_hash:
                return current_text
    return None
\end{lstlisting}

\subsection{ Quantum Implementation}
\subsubsection{State Preparation}
State preparation is an essential process in quantum computing that involves setting up the quantum system to the desired state before performing any quantum operations. When a quantum circuit is initialized, all qubits are set to the state {|0⟩}. In a typical state preparation step, qubits are changed to be in the desired state, which generally is a superposition of states. In QIris, the state preparation involves setting the search space in which Grover’s Algorithm is applied. 

Assume that in a 4-qubit quantum circuit, the potential measurement outcomes range from binary 0000 to 1111, which correspond to integers 0 to 15. If there are only 7 items (1, 2, 8, 9, 10, 12, 15) in the search space (array elements), the probability of measuring the other 8 items will be 0. Therefore, in this case, the state preparation will be configured to ensure that the measurement outcomes are limited to these 7 items only. 

To set the state preparation for the above example, an array {[}0, 1, 1, 0, 0, 0, 0, 0, 1, 1, 1, 0, 1, 0, 0, 1{]} with the positions 1, 2, 8, 9, 10, 12, 15 marked to be part of the superposition representation. The Frobenius norm \cite{sd_frobenius} of the array is then calculated. Subsequently, the elements in the array are converted to probability amplitude values through the division of each element with the calculated Frobenius norm. The superposition of for the above array can be expressed as: 

\[
\frac{1}{\sqrt{7}} |1\rangle + \frac{1}{\sqrt{7}} |2\rangle + \frac{1}{\sqrt{7}} |8\rangle + \frac{1}{\sqrt{7}} |9\rangle + \frac{1}{\sqrt{7}} |10\rangle + \frac{1}{\sqrt{7}} |12\rangle + \frac{1}{\sqrt{7}} |15\rangle
\]

To derive \(\frac{1}{\sqrt{7}}\) as the probability amplitude, the Frobenius norm for the array is calculated to be  \(\sqrt{7}\). Therefore, each element that has a non-zero value in the array is divided by the Frobenius norm to create the final array:

\[
[0, \frac{1}{\sqrt{7}}, \frac{1}{\sqrt{7}}, 0, 0, 0, 0, 0, \frac{1}{\sqrt{7}}, \frac{1}{\sqrt{7}}, \frac{1}{\sqrt{7}}, 0, \frac{1}{\sqrt{7}}, 0, 0, \frac{1}{\sqrt{7}}]
\]\

This array is then passed to the Statevector function to generate the state preparation circuit. The concept above is utilized in QIris during the state preparation process. The code used to generate the state preparation is shown in Listing \ref{lst:stateprep}. 

\begin{lstlisting}[caption={State Preparation},captionpos=b,label={lst:stateprep}]
norm = np.linalg.norm(data)
data = data / norm
state = Statevector(data)
\end{lstlisting}

\subsubsection{Grover’s Algorithm}
Grover’s Algorithm comprises two main components: the oracle and the amplitude amplification. The oracle marks the computational basis state as the state that it is interested in finding, while the amplitude amplification increases the amplitude of the marked state \cite{8622457}. 

The oracle generate function is implemented to generate the oracle for the Grover’s Algorithm based on the target hash lookup value. The input is first converted to binary representation in little endian format before being transformed into a boolean expression. The boolean expression is then passed into the PhaseOracle function from qiskit.circuit.library, which will generate the oracle circuit. The code for the oracle circuit is shown in Listing \ref{lst:oragen}.  

\begin{lstlisting}[caption={Oracle Generation},captionpos=b,label={lst:oragen}]
def oracle_generate(target_state):
    binary_string = format(target_state, '04b')[::-1]
    boolean_expression = ""
    for x in range(len(binary_string)):
        if binary_string[x] == "0":
            boolean_expression += f"~x{x} & "
        else:
        	boolean_expression += f"x{x} & "
    boolean_string = boolean_expression[:-3]
    return PhaseOracle(boolean_string)
\end{lstlisting}

The algorithm then uses an amplification circuit, also known as the diffusion operator, which amplifies the probability of measuring the correct state. The process involves repeatedly applying the oracle and the amplifier a certain number of times, which is determined by the formula: 

\begin{equation}
    \centering
        \left\lfloor \frac{\pi}{4} \times \sqrt{N} \right\rfloor
    \label{eq:captioned_equation}
\end{equation}
    \text{where \(N\) equals the number of elements in the search space}

The GroverAlgo function is implemented to generate the complete Grover’s algorithm circuit and run the circuit through the simulator. Firstly, the Grover operation circuit is generated by passing through the oracle and state preparation circuit to the GroverOperator function. This Grover circuit is then appended to the end of the state preparation, with the number of iterations determined based on the formula \eqref{eq:captioned_equation} for the optimal number of iterations. At the end of the circuit, all qubits are marked for measurement. The final Grover’s algorithm circuit is then transpiled and send to the simulator. 

\begin{lstlisting}[caption={Grover's Algorithm Function},captionpos=b]
def GroverAlgo(init_state, oracle, iter, bits):
    grover = GroverOperator(oracle, state_preparation=init_state)
    for _ in range(iter):
        init_state.append(grover, range(bits))
    init_state.measure_all()
    backend = Aer.get_backend('qasm_simulator')
    compiled_circuit = transpile(init_state, backend)
    job = backend.run(compiled_circuit)
    counts = job.result().get_counts()
    return counts
\end{lstlisting}

To determine whether the Grover’s Algorithm has successfully found the target state, we evaluate the result based on the measurement counts. Specifically, the state with the highest count must represent more than 50\%\ of the total shots. In QIris, the total number of shots is set to 1024, so the maximum count for a state must exceed 512 for the algorithm to be considered successful. 



\section{FUTURE WORKS} 
With this Quantum Rainbow Table, QIris demonstrates a hybrid approach that integrates classical linear search with quantum computing techniques. Moving forward, several avenues for future work are envisioned. Firstly, exploring and implementing the use of other hash functions, such as SHA-1 and SHA-2, to evaluate the versatility and performance associated with different hashing algorithms. The main focus will be on the speed of quantum computers as well as the hashing process, emphasizing the time efficiency of classical and quantum hashing. Furthermore, this project opens the possibility of exploring concepts like quantum salt, where the salt value is placed in a superposition state, which could potentially lead to significant improvements in the security and efficiency of hashing processes. Lastly, more studies will be required to investigate the feasibility of these innovations and assess various ways to optimize both classical and quantum hashing techniques. This research direction could pave the way for more advanced cryptographic methods and applications, leveraging the full potential of quantum computing.

\section{CONCLUSION}
Our work explores the forefront of integrating quantum computing with classical techniques to enhance the rainbow table, culminating in the development of QIris, a Quantum Rainbow Table. By utilizing Grover's algorithm, QIris operates with a search time complexity of $O(\sqrt{N})$, as opposed to the classical rainbow table’s $O(N)$, thereby demonstrating the computational advantages of quantum computing. This hybrid approach effectively combines classical and quantum methods, overcoming hardware limitations. We anticipate that our implementation of the quantum rainbow table will serve as a valuable tool for researchers to further investigate quantum-based hash attacks.
In addition, our research underscores the potential of quantum computing in cybersecurity, encompassing a variety of projects outlined in future works. QIris represents a novel application of quantum computing to cryptographic analysis, with the expectation that it will open new avenues for more secure systems as quantum hardware continues to evolve.

\end{document}